# The Elephant in the Room: Why Transformative Education Must Address the Problem of Endless Exponential Economic Growth

CHIRAG DHARA[1], VANDANA SINGH[2]

[1]*Indian Institute of Tropical Meteorology (Ministry of Earth Sciences), Pune, India*
[2]*Framingham State University, Framingham, USA*

**Abstract**

A transformative approach to addressing complex social-environmental problems warrants reexamining our most fundamental assumptions about sustainability and progress, including the entrenched imperative for limitless economic growth. Our global resource footprint has grown in lock-step with GDP since the industrial revolution, spawning the climate and ecological crises. Faith that technology will eventually decouple resource use from GDP growth is pervasive, despite there being practically no empirical evidence of decoupling in any country. We argue that complete long-term decoupling is, in fact, well-nigh impossible for fundamental physical, mathematical, logical, pragmatic and behavioural reasons. We suggest that a crucial first step toward a transformative education is to acknowledge this incompatibility, and provide examples of where and how our arguments may be incorporated in education. More broadly, we propose that foregrounding SDG 12 with a functional definition of sustainability, and educating and upskilling students to this end, must be a necessary minimum goal of any transformative approach to sustainability education. Our aim is to provide a conceptual scaffolding around which learning frameworks may be developed to make room for diverse alternative paths to truly sustainable social-ecological cultures.

## Introduction

A major roadblock to effective climate change education is the lack of a radical vision in the global educational community (Kwauk, 2020). "Transformative education" is increasingly being recognized by UNESCO and other educational organizations as central to realizing a sustainable future (Odell et al., 2020). Recognized to a lesser extent is the need for transdisciplinarity in climate education (Singh, 2020) since social-environmental problems tend to transcend disciplinary boundaries.

A key aspect of transformational learning is to "foster deep engagement with and reflection on our taken-for-granted ways of viewing the world, resulting in fundamental shifts in how we see and understand ourselves and our relationship with the world." (Journal of Transformative Education, n.d.)

Researchers suggest that this "fundamental shift" can occur through a transformation-oriented educational approach, including what Mezirow and Taylor (2009) refer to as a disorienting dilemma. The work of Bain (2004) and others support this idea, suggesting that a necessary element of a "natural critical learning environment" is to present students with experiences that violate their existing paradigms, as a first step toward constructing a new mental image of the world.

This approach takes on particular significance because immersion in a paradigm without being conscious of that immersion is a kind of blindness that can prevent us from acknowledging the falsity of some of our unexamined, underlying assumptions (Singh, 2020). In the larger context of critical global crises of anthropogenic origin (IPCC, 2018; IPBES, 2019), the result of such blindness may prevent us from taking the needed actions toward a truly sustainable future.

A crucial aspect of paradigm blindness is the persistence of certain ideas and assumptions that become epistemological roadblocks to both effective education and effective action. These assumptions can lead to contradictions between our intentions and our actions. We discuss a serious contradiction that has been identified within the UN Sustainable Development Goals (SDGs) (Hickel, 2019), and suggest that our elucidation affords an opportunity – in the classroom and beyond – to realize the deepest order of change/ learning in the hierarchies of learning described by scholars of transformative education (Sterling, 2011), namely, "epistemic learning," which can potentially lead to the paradigm change necessary for truly sustainable human–natural futures on our planet.

The persistent idea that we focus on here is the notion of *limitless exponential economic growth*. The size of the economy, usually measured by Gross Domestic Product (GDP), on per capita basis, is found to be a useful proxy for progress, being closely correlated with a host of indicators such as life expectancy (Roser, Ortiz-Ospina, et al., 2013), (reduced) child mortality, (Roser, Ritchie, et al., 2013), and average years of schooling (Our World in Data, 2017). Hence the idea of limitless economic growth as essential to our future has persisted in both the literature and the media, often alongside discussions of sustainable development (including in UNESCO documents, UNESCO, 2016; Odell et al., 2020). Yet, the high average levels of affluence in the Global North, born of uncontrolled GDP growth, and the small but super-affluent classes within the Global South, born of unequal wealth distribution, are both linked to disproportionately high levels of material consumption.[1] Economic growth without limits has become an unconditional imperative (Richters & Siemoneit, 2019) for the entrenched socioeconomic system. While a certain level of GDP seems to be indicated for human well-being, the notion of economic growth has no sufficiency clause. This lack of a limit raises the critical question of whether future developments, such as in technology, may eventually be able to reconcile endless economic growth with long term sustainability. We grapple with this question in the following way. First we point out the connection between high material consumption and our current environmental crises, including climate change and loss of biodiversity. On the basis of this analysis, we propose a functional definition of sustainability. Third, and centrally, we then argue that the contradiction between endless exponential economic growth and genuine sustainability is serious and that it is effectively irreconcilable on physical, mathematical, and logical grounds. Since the eradication of poverty and inequality is also crucial, we allude briefly to some of the proposals in the scientific literature reconciling human well-being and sustainability.

Many scholars have pointed out the problematic implications of endless economic growth (McBain & Alsamawi, 2014; Hickel & Kallis, 2019; Malik et al., 2019); however neither have the various arguments been elucidated nor has their application to the educational domain been made to our satisfaction. In this context we identify a necessary minimum goal that must underpin any transformative approach to education for sustainability, and we provide examples of where and how these may be incorporated (concluding section and Table 1). Our intent in this chapter is not to present a lesson plan or framework, but to provide the conceptual scaffolding for educators and concerned citizens to develop their own learning frameworks around this central contradiction.

### Sustainability From a Resource Perspective

All aspects of industrialized societies, ranging from food production and consumption, building and infrastructure construction, power production, modes of transport and communication, among others, make intensive use of primary and derived physical resources, such as land, water, cement, plastics, glass, rubber, fossil fuels, and metals, including rare earths. Thus, the remarkable technological advancements of the modern era have been enabled and sustained by an exponentially rising[2] rate of resource extraction and materials production over the past two centuries (Figure 1).

The industrial-scale extraction of raw materials has come about through large-scale mining operations, often in remote and biodiverse areas around the world. These raw materials are used in constructing and maintaining such large-scale infrastructure projects as roads, rail, and power plants, and in the manufacture of various products for industrial, agricultural, and personal use.[3] These products and projects eventually reach the ends of their lives and are scrapped, even though some go through a few cycles of reuse or recycling.

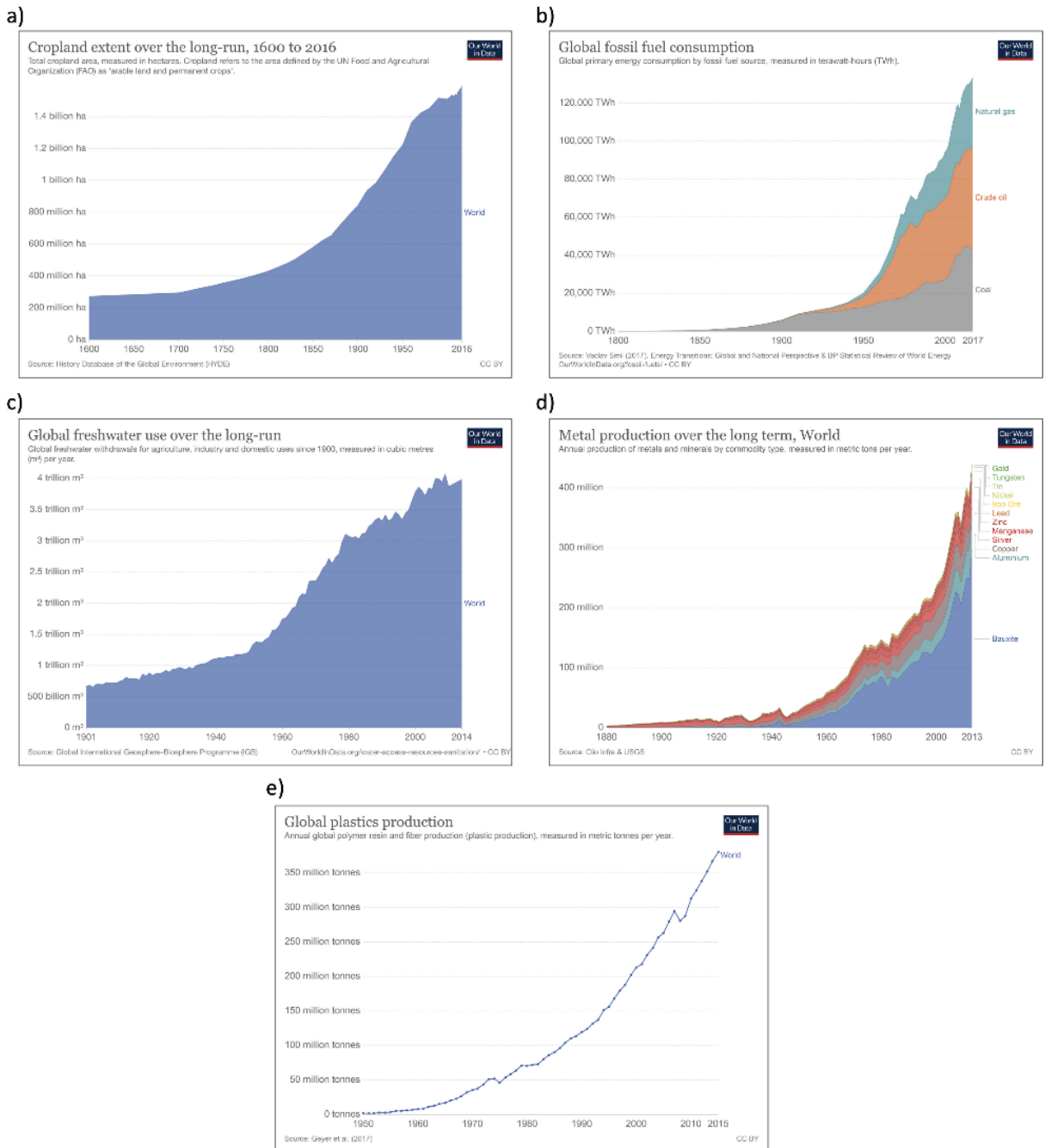

FIGURE 1    Industrial era exponential rise in the use of primary and derived physical resources: (a) cropland (Ritchie & Roser, 2013), (b) fossil fuels (Ritchie, 2017a), (c) fresh water (Ritchie, 2017b), (d) metals (Our World in Data, n.d.), and (e) plastic (Ritchie, 2018).

Each of the stages from extraction through disposal requires the use of land, water, and energy, for which humans have invaded and destroyed natural habitats (Figure 2), and thus provoked our ongoing ecological crisis, the sixth wildlife extinction event[4] in Earth's history (Ceballos et al., 2017).

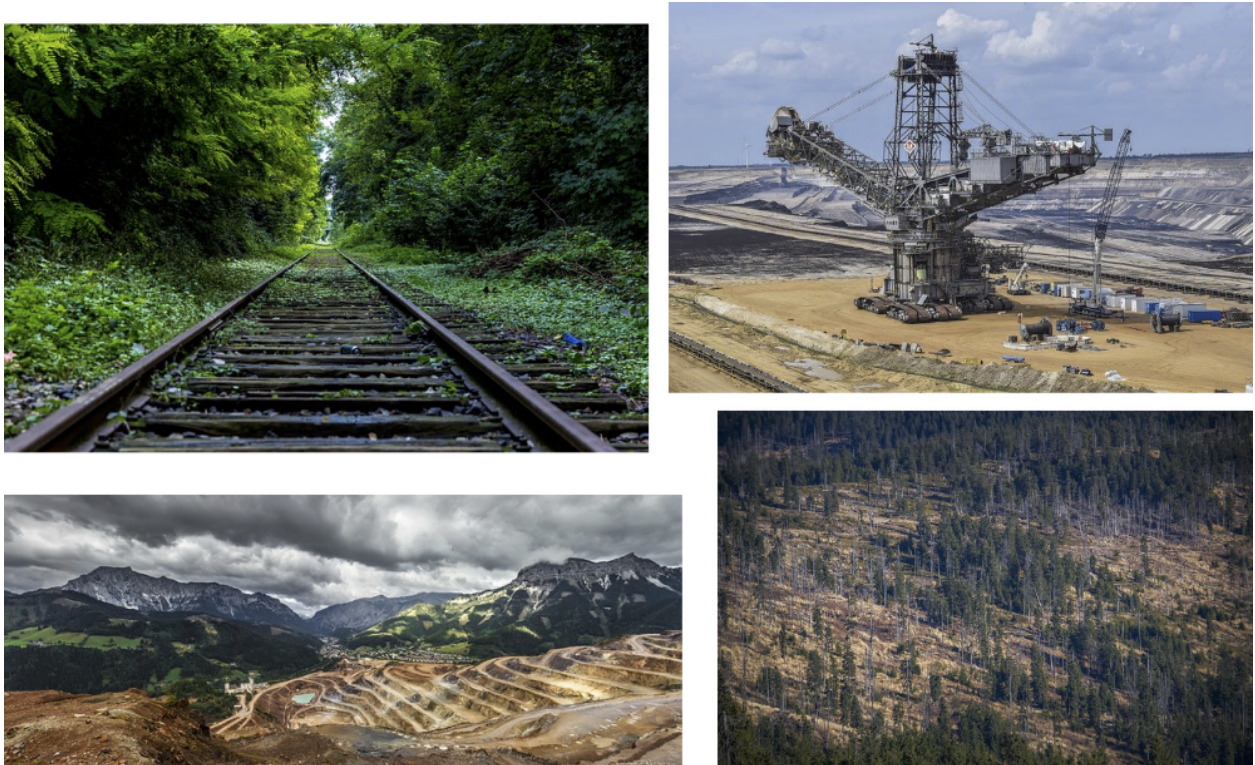

FIGURE 2      Ecological destruction for resource extraction and infrastructure construction (Pixabay images).

Concurrently, each of these stages causes large-scale air, soil, and water pollution, with the attendant consequences for human and natural health (Figures 3 and 4). In particular, the carbon pollution from fossil fuel use is the key cause of the escalating climate crisis.

In other words, the global climate and ecological crises are both symptoms of our prolific use of finite planetary resources.

a)

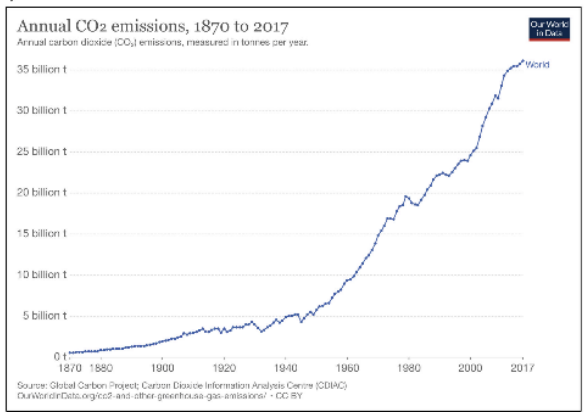

b)

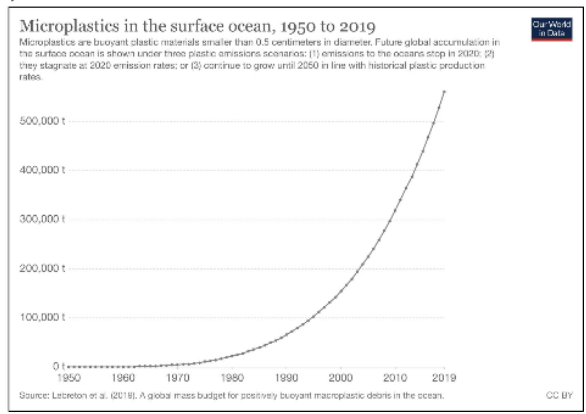

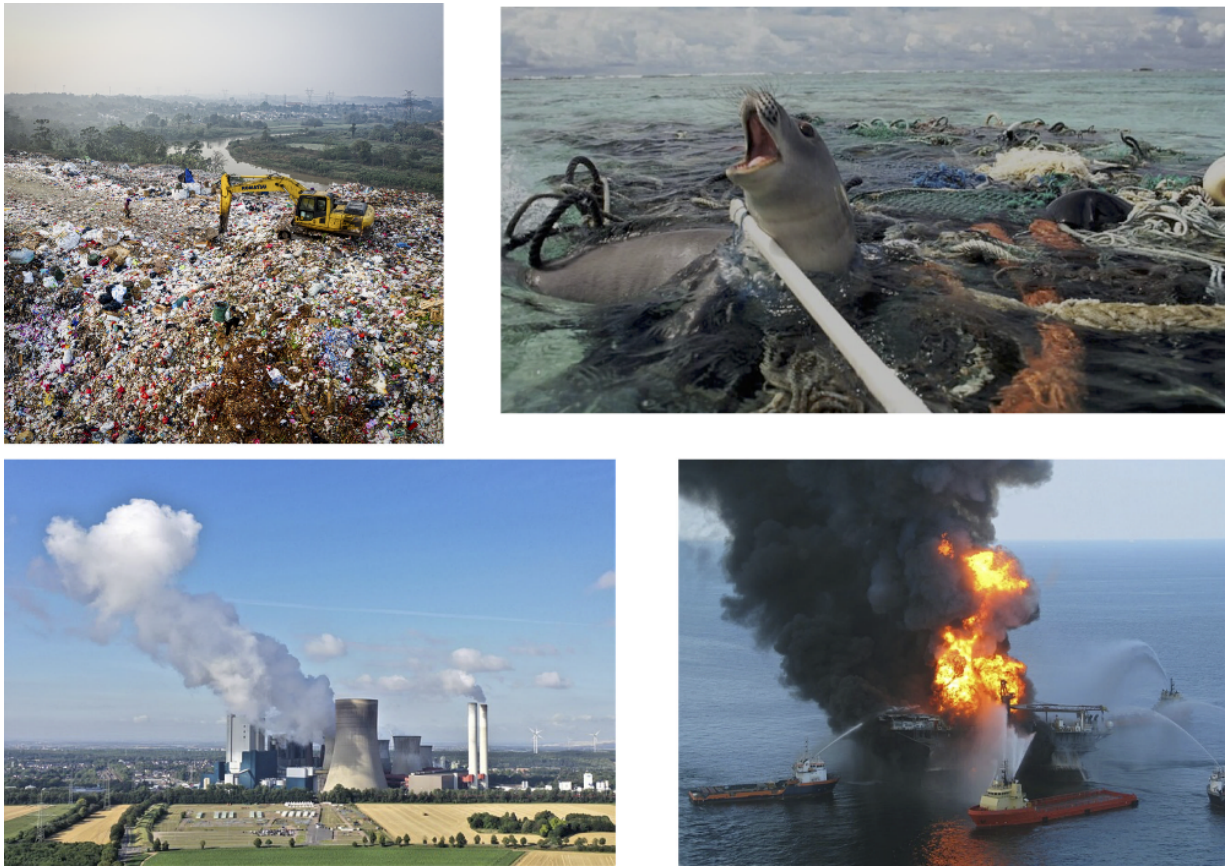

FIGURE 4     Water, land. and air pollution (Credit for top right image: Nels Israelson under CC BY-NC 2.0. Others: Pixabay images).

A metric called the material footprint (MF) quantifies the rate at which humans are expropriating physical resources from nature (Wiedmann et al., 2015). MF aggregates the total mass of construction minerals, biomass, fossil fuels, and metal ores at country and global levels to give a snapshot of our burden the planet. The global material footprint increased from 54 billion metric tonnes in 2000 to 92 billion metric tonnes in 2017, an increase of 70% in a mere 17 years (UN Statistics Division, n.d.).

The resource perspective reveals not just the unsustainability of the global resource consumption but also leads to a natural minimum condition for transition toward genuine long-term sustainability, namely, *all resource use curves must be simultaneously flatlined, and pollution curves must be extinguished*[5] (Figure 5). This formulation constitutes a more useful and practical definition of sustainability than "meeting the needs of the present without compromising the ability of future generations to meet their own needs" (Brundtland, 1987). For the remainder of this chapter, we adopt this practical definition of *genuine long-term sustainability* as a restatement of SDG 12.

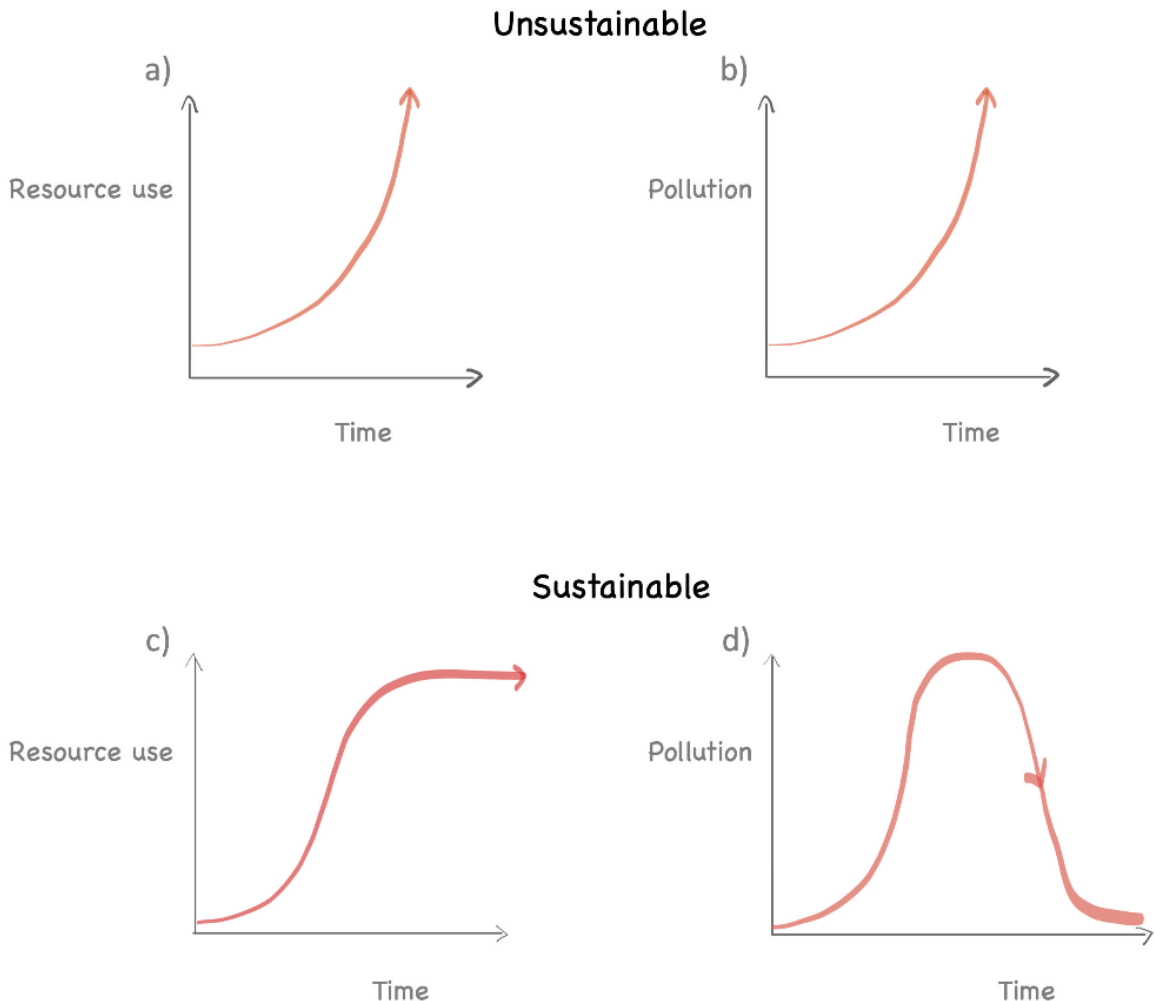

FIGURE 5          Sustainability from a resource perspective. Exponentially rising resource use and pollution, represented by (a) and (b), are unsustainable. We define sustainability as flatlined resource use as in (c), and extinguished pollution as in (d). (Credit: Aditi Deshpande).

The grave consequences of the climate and ecological crises to life on the planet make it vital that sustainability in the sense of Figure 5 must be foregrounded in SDG 4.7. In particular, we argue that generating and widely instilling the pertinent knowledge and skills must be a necessary minimum goal of SDG 4.7 (concluding section and Table 1).

How do we flatten multiple exponentially-rising resource-use curves simultaneously? It is of critical importance that we view this systemic problem from a systems perspective, and that we ask what the fundamental reason(s) is (are) that has (have) necessitated unsustainable growth in resource use in the industrial era.

### Exponential Economic Growth

Let us turn our attention to the dominant grow-or-bust economic doctrine that stipulates that an economy is healthy only if it grows by a certain percentage every year. What may not be immediately obvious is that "percentage growth" amounts to exponential growth. At their roughly 2% growth rate, the economies of the Global North countries double every 35 years. India's economy would double every 10 to 12 years if it sustained growth at the generally touted rate of 6 to 7%.

The problem is that rising wealth associated with economic growth is linked not only to meeting basic human needs, but also to ballooning luxury consumption: electronic devices, air-conditioners, private vehicles, flights, cruise ships, house furniture and appliances, junk food, and fashion, to name just some. The manufacture, the transport, and the creation of the means of consumption of these are intimately linked to the use of physical resources. Thus, we would expect material growth, and pollution, to grow in lockstep with economic growth, which is exactly what has occurred, as seen in Figures 1, 3, and 6.

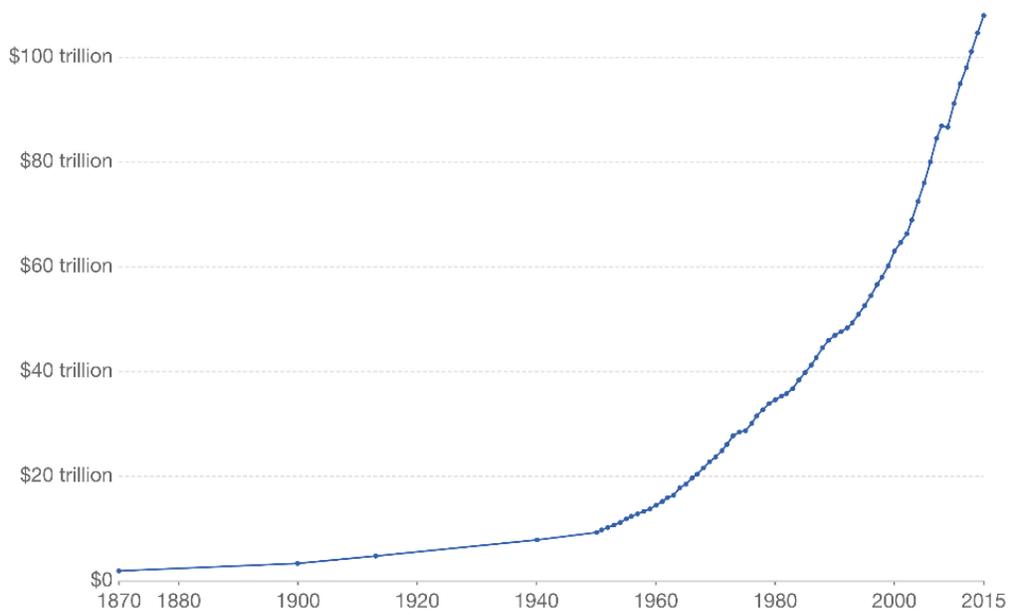

FIGURE 6    Global GDP growth in the industrial period (Roser, 2013).

Closer examination reveals an unmistakably sharp rise in materials use and carbon pollution post World War II, exactly in sync with the sharp rise in GDP. The foregoing arguments lay out the physical basis for why these are not mere fortuitous correlations, but instead causal associations.

### A Fundamental Conflict Among the SDGs

SDG 12 (Responsible Consumption and Production) calls for stabilizing our use of planetary resources, consistent with our observations above. Yet, SDG 8 (Decent Work and Economic Growth) calls on countries to promote sustained economic growth, although with a "sustainable" and "inclusive" character (United Nations, 2015).

In the following sections, we argue that despite claims to the contrary, SDGs 8 and 12 are very likely irreconcilable. This inconsistency may pose a major barrier to the effectiveness of SDG 4.7, which seeks to "ensure that all learners acquire the knowledge and skills needed to promote sustainable development, including, among others, through education for sustainable development and sustainable lifestyles," if this contradiction remains unacknowledged and unaddressed. We discuss this point in the concluding section.

### *The Imperative for "Decoupling" Resource Use From Economic Growth*

The conventional response to the problem of reconciling indefinite growth with reductions in material use[6] is to "decouple" economic growth from the use of physical resources. This is the fundamental premise of concepts such as the "circular economy" (see for example Ellen Macarthur Foundation, 2017), "green/sustainable growth" (SDG Knowledge Platform, n.d.), and the Green New Deals recently proposed by progressive movements in the United States (Friends of Bernie Sanders, n.d.) and the UK (Labour for a Green New Deal, n.d.).

Decoupling is required not only with immediate effect (IPCC, 2018), but  also to be indefinitely sustained. That is to say, even if the global economy increases 50- or 100-fold over that of the present day, the physical resources circulating in the economy must be no more than those circulating at present. In fact, the quantity must be lower, since we have already severely compromised biospheric integrity (IPBES, 2019).

But can decoupling be achieved across the board, and at scale? Can we sustain it indefinitely? Is there historical precedent that we can draw from?

The formulation of the SDGs and Green New Deals are predicated on the presumption that the answers to each of these questions is an unequivocal yes, based on an implicit and largely unquestioned faith that technological innovation will be our deliverance from these crises and that economic growth may continue indefinitely. Major solutions usually proposed to decoupling materials use and pollution from economic growth include transitioning to renewables, improving the energy efficiency of appliances, recycling and reducing waste, and expanding digital use.

In the following section, we argue that strong mathematical, physical, logical, pragmatic, and behavioral constraints serve to limit technology's ability to deliver long-term sustainability. In doing so, we question the scientific and evidential foundations of the SDG formulation. Let us examine these in more detail.

### *Mathematical Constraints: Exponential Growth*

Exponential growth can be understood as a rate of rise in which the "doubling time" is constant, a concept whose ramifications are best understood through a thought experiment.

Let us imagine that the currently known stock of fossil fuels is calculated to last for 100 years, with demand growing at 2% per year. Suppose we discover substantial new reserves tomorrow that immediately raises the stock to four times that amount. How much longer will the enlarged stock last, assuming all else remains constant?

While it may be tempting to think that we would be covered for 400 years, in reality, at demand that grows at 2% per year, this vastly increased stock will last only 170 years. This result is the consequence of demand's doubling every 35 years at its 2% rate of growth.

What this kind of calculation means is that material use can be flattened if, and only if, the discovery of new stocks of *all* physical resources, or the improvement in the efficiency of their use, proceeds indefinitely at the same, or greater, exponential rate as GDP growth.

Resource stocks are, however, necessarily finite on a finite planet. In addition, while improving the efficiency of resource use is often seen as a major contributor toward achieving sustainability, efficiency improvements have hard upper limits, as we discuss below.

### Physical Limits to Efficiency

In September 2017, the Formula 1 car company, Mercedes, announced (Gilboy, 2017) that their engine had achieved a "thermal efficiency" exceeding 50%, meaning that the engine was converting more than 50% of the energy of the fuel into useful work to power the car. This was remarkable, since most such engines usually operate at only 20 to 40% efficiency (Office of Energy Efficiency, n.d.).[7] The significance is that the amount of fuel used to power a car with an engine efficiency of 25% can power *two* cars that have double the engine efficiency, thus decoupling fuel (resource) use from the growing demand for cars. Yet there is a hard upper limit to how efficient a car engine can become. The laws of thermodynamics[8] guarantee that no such engine can ever become more than 80% efficient.[9] In fact, in practice the Mercedes engine would struggle to exceed even 60% efficiency, which is around the highest ever achieved.[10]

Other pertinent examples include the physical limit of about 45% on the efficiency of photovoltaic cells (Do the Math, 2011), and a 1 W/m$^2$ energy generation capacity limit on large scale wind power installations (Miller et al., 2015).

The point of these arguments is that while improvements in efficiency can deliver short-term decoupling from demand, efficiency is limited by physics, and at the same time, there is *no sufficiency limit on demand.* Once peak efficiency is achieved – and many of our technologies are operating close to those limits – further increase in demand will necessarily drive an increase in resource use.[11] Additionally it becomes less time-and-cost-effective to invest in efficiency improvements when we approach physical limits, because return on investment declines under those conditions.

### Pragmatic Limits

Waste in global food production is estimated at nearly 30% (FAO, n.d.). A substantial quantity of fresh water is wasted during transport and usage (see, for example, Dharma Rao, n.d.). A rapid increase in agricultural productivity coupled with reduction in wastage may undoubtedly allow demand to rise without a corresponding increase in the global land and water footprint, with a resultant short-term decoupling.

Yet, the logical consequence of a complete elimination of wastage is also the elimination of scope for further improvement: a pragmatic limit to improving the efficiency of use of a physical resource. Further rise in demand (such as land used for agriculture, transport infrastructure, server farms, solar power plants, and so on) will necessarily "recouple" resource use with demand. And demand is indeed projected to continue rising. Studies have found that the footprint of affluent nations on land and ocean grew in size by 30% each time their income doubled (Weinzettel et al., 2013).

### Logical Constraints

The concept of the "circular economy" has been at the forefront of global discussions on decoupling material use from the demands of an exponentially growing economy. (Ellen Macarthur Foundation, 2017; UNEP, 2019; World Economic Forum, n.d.; OECD, 2020; European Commission, 2020). The means intended to achieve it are a combination of effectively recycling or "upcycling" all end products to feed back into the system so that few or no virgin materials are necessary.

However, the very premise of "circularity" contains a logical flaw: even complete recycling of all material resources of the previous year can in no way meet the requirements of the current year if the demand for resources has grown (Figure 1). At best it can meet the demand for resource use only to last year's level. Added demand this year will necessarily require virgin materials.

### Limits to Recycling

Glass and metals can be recycled almost indefinitely without loss of quality (Holmes, 2017). However, some materials can be recycled only a limited number of times before becoming too degraded to be further recycled (Howard, 2018). Plastics and paper, for instance, become unrecyclable within about five recycling cycles.

In addition, recycling techniques are non-trivial and material specific (Recycling Centers, 2020), realities that pose pragmatic difficulties. For instance, electronic devices consist of a multitude of components incorporating varying materials. Let us consider an iPhone as an example. It is made up of several metals and other elements (von Kessel, 2017). Some of the procedures necessary for a complete recycling of the total stock of iPhones would be as follows:

- Building, maintaining, and constantly expanding the capacity of recycling centers in order to handle the growing volume of spent phones.
- Recovering used phones from around the world.
- Dismantling various components, such as the screen, processor, and so on, and separating out the constituent raw materials, such as silicon, copper, aluminum, and silver (Figure 7).
- Shipping to specialized recycling centers for separate recycling of each raw material.

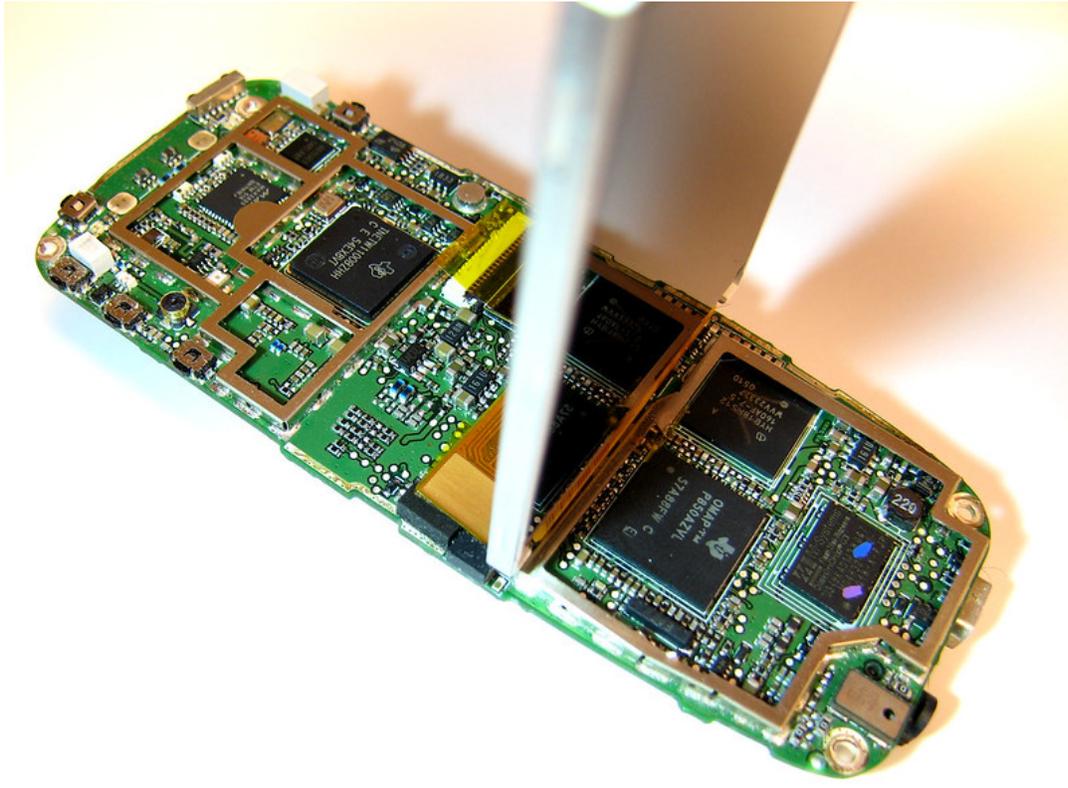

FIGURE 7    Raw materials constituting a microprocessor chip are part of a "diffused whole" and need to be segregated and cleaned to allow separate recycling of each (Andrew Magill).

Each of these steps must be performed with high efficiency in order to achieve a tolerable recycling fraction. In addition, these steps require expenditure of physical resources, such as energy for segregating raw materials and water for cleaning. Thus, aside from the pragmatic difficulty of achieving each step with nearly 100% efficiency, the recycling process itself necessitates the use of a significant amount of additional material resources.

### Behavioral Traits and Problem-Shifting

One might expect that having a new energy-efficient television would reduce electricity use. Yet empirical studies have found that potential gains may be offset in part because, all else remaining the same, lowered electric bills may encourage people to buy larger, or multiple, television sets. This type of rebound effect has repeatedly frustrated attempts at resource conservation through efficiency gains (Druckman et al., 2011; Giampietro & Mayumi, 2018). This phenomenon brings us to the issue of behavioral constraints.

A useful thought experiment here is to imagine the "perfect electric car": solar powered, efficient, reliable, affordable. What happens next?

Buying such a car would engender little guilt. Everyone could buy one, and could drive longer distances, since ostensibly neither energy nor pollution is at issue. Having such a car might also act as a disincentive to using public transport.

However, the approximately metric tonne weight of the car constitutes raw materials ranging from metals to glass, plastic, and rubber. A sustained global spike in the demand for such a car would drive increased demand for these raw materials, exacting environmental costs at every stage:
– Manufacture: pollution from increased mining for raw materials.
– Use: pollution from constructing or expanding infrastructure such as roads and bridges, especially if fragile ecosystems are disrupted.
– Disposal: pollution from raw materials that cannot be recycled or biodegraded.

In short, such an invention runs the risk of becoming an elaborate exercise in *problem-shifting*.

### *Lack of Historical Precedent and the Speculative Potential of Future Technology*

Finally, we argue that even the evidential foundations of the SDG goals are seriously in doubt, since practically no empirical evidence exists in any country of any genuine decoupling.

According to OECD (2011), G8 countries halved their resource intensity between 1980 and 2008, and Canada, Germany, Italy, and Japan succeeded in decoupling their *domestic materials consumption*[12] from economic growth in absolute terms. However, Wiedmann et al. (2015) performed a careful accounting of the Material Footprint (MF)[13], including those embedded in internationally traded products, and reported that

> the MF has kept pace with increases in GDP and no improvements in resource productivity at all are observed when measured as the GDP/MF. This means that no decoupling has taken place over the past two decades for this group of developed countries.

Basically, the Global North has offshored a substantial part of its production, and thus also the associated emissions (Davis & Caldeira, 2010) and ecological destruction, to the developing countries of the Global South.

The ongoing digital transformation – nanotechnology, biotechnology, artificial intelligence, Internet of Things – is often touted as the enabler of absolute decoupling in the near future through efficiency gains (Ekholm & Rockström, 2019). Yet, such claims suffer from several deficiencies, such as implicitly considering only short-term decoupling,[14] not accounting for all the limits discussed herein, and being replete with speculative language on the future potential of technology. In fact, thus far, digitalisation has increased consumption and remained coupled with the indirect use of energy and materials (Parrique et al., 2019; Wiedmann et al., 2020).

### Discussion and Recommendations for Education

The arguments we have set forth must not be construed as suggesting that technology has no place in moving toward sustainability. Indeed, we urgently need to transition to *more circular* economies with the use of renewables, improving efficiency to the highest possible extent, reducing waste, recycling and upcycling, and employing other such initiatives (Hawken, 2017). The interpretation of the constraints we have elucidated here is that technology *alone* is highly unlikely to engender genuine long-term sustainability, and unquestioned faith in the current growth paradigm amounts to staking our collective survival on belief rather than on science or on evidence. The fundamental problem is incontrovertibly overconsumption by the affluent and the unconditional growth paradigm (Wiedmann et al., 2020). But if we abandon the current growth paradigm, which, as mentioned earlier, correlates with certain development indicators, what about human well-being?

The current default world view appears to be that economic growth is the *only* way to achieve economic justice and human well-being. Aside from the environmental problems discussed earlier, this world view also ignores the violence associated with economic growth: land grab and dispossession of people, especially Indigenous peoples, and unprecedented increase in a form of social inequality that puts enormous power in the hands of those who benefit most from the current system. In addition, the relationship between the economic growth model and the increase in development indicators, on one hand, and the creation of poverty and inequality, on the other hand, is complex (Harris-White, 2006, Coffey et al., 2020). What, then, are the solutions?

Many alternatives to the current system have been proposed, ranging from reformist to radical, such as steady-state economics, degrowth, "agrowth," eco-anarchism, cross-pollinations, and variations thereof (Daly & Farley, 2003; Jackson, 2009; Daly, 2014; van den Bergh, 2017; Victor, 2019; Alexander, 2015; Smith, 2016; Kallis, 2018; Alexander & Gleeson, 2019; Nelson & Timmermans, 2011). These approaches seek to decouple human well-being from GDP growth, including a strong focus on sufficiency, equity, cooperation, and social justice. Goals include individual down-shifting among the affluent classes, decentralized production, constant monitoring of human and planetary well-being, basic income and job guarantees, setting maximum income levels, changing lifestyles and cultures through grassroots action, stronger regulation of ecologically destructive industries, and eco-villages. In addition, multiple local, grassroots experiments in alternatives are being practiced by communities around the world, particularly the Global South (Kothari et al., 2019; Gerber & Raina, 2018). A detailed discussion of these is, however, outside the scope of this chapter.

Our clarification of the conflict between SDG 8 and 12 makes the case for an urgent need to seriously consider alternatives to economic growth that reconcile human well-being and a sustainable future. To this end, we propose the following necessary minimum goal of a transformative approach to sustainability education:

> That SDG 12, *centered on our functional definition of sustainability,* be foregrounded in education. The conflict between SDGs 8 and 12 must be emphasized in line with the arguments set forth here, as a portal to thinking about alternatives that embrace both human well-being and long-term sustainability. This, then, allows SDG 4.7 to become a truly useful tool toward genuine sustainability.

Consistent with SDG 4's reference to "lifelong learning," effective sustainability education should not be limited to students, but rather extended to citizens, policymakers, and government and corporate leaders, who perhaps more than others need a paradigm shift.

TABLE 1     Some alignments between our arguments and topics taught in high school and college.

| Key concepts | Intersections with topics and disciplines |
| --- | --- |
| Material consumption | Students can examine data on expansions in material footprints, the role of the affluent in cementing the paradigm of economic growth, and their impact on ecosystems and communities. They can examine how individuals and communities are influenced by this paradigm and discuss possible systemic and personal shifts away from increasing material consumption so as to "flatten the curve." |
| Exponential growth | Study of exponential growth need not be limited to math classes; as a |

| and mathematical limits | pervasive global phenomenon, whether in the context of pandemics or GDP or plastic pollution, it can be taught and contrasted with default linear thinking in classes from the sciences to economics, history, and the social sciences. |
| --- | --- |
| Physical limits to efficiency | While thermodynamics is taught in most physical science classes in high school and college, applying its laws to efficiency limits of technology is not the norm; doing so would drive home the idea that technofixes alone will not solve our complex problems. It is crucial that the essential arguments also be taught in multiple disciplines outside the sciences. |
| Pragmatic limits to reducing waste | Students usually study the seriousness of issues like food, water, and waste. "Reduce, reuse, and recycle" is the common refrain in environmental education and international campaigns. While we must strive to achieve all three Rs, educational systems must go even further and emphasize yet another "R" – "refuse." |
| Limits to recycling | Most educational efforts to inculcate pro-environmental behaviors in children and youth tend to highlight recycling as the go-to behavioral change of choice. While there is scope for considerable improvement in recycling rates, extolling the virtues of recycling without a comprehensive discussion of the inherent limits poses the danger of its being easily co-opted to justify incremental – and ultimately inadequate – changes to the status quo instead of aspiring towards truly transformative change. |
| Behavioral traits and problem-shifting | Studying behaviors and problem-shifting gives students an opportunity to explore ethical and psychological dimensions of the value system of the dominant socioeconomic paradigm in the context of sustainability. |
| Lack of historical precedent for decoupling | Documenting scant decoupling success is relatively new, research that has not, to our knowledge, been integrated into curricula in high schools and colleges. It can be explored along with material consumption (above) to drive home the point that despite claims to the contrary, no nation has achieved decoupling; moreover this calls upon us to consider and connect *both* national and global footprints. |

In Table 1 we suggest ways in which different aspects of our argument may align with existing education topics and disciplines in high school and college. However it is not our intent to advocate for a reductionist, piecemeal approach. The power of transformational learning, coupled with the horizon-expanding, systems approach of transdisciplinarity, potentially allows for an *epistemic shift* within the learner – an irreversible cognitive and affective shift in perspective that can potentially lead to the recognition of paradigm blindness and the emergence of new paradigms. Within such a broad framework, we urge educators to employ high-impact practices such as project-based learning (English & Kitsantas, 2013), social and emotional learning (Weissberg et al., 2015), and case-based learning (Yale Poorvu Center, 2020) that can guide students toward discovering and exploring the contradiction between SDGs 8 and 12. Examining this incompatibility provides the opportunity to make invisible paradigms visible and contestable, and opens space for considering other models of social-economic-ecological relationships that seek to promote both human well-being and ecological harmony.

We maintain that yet another "orienting anchor" for any meaningful implementation of SDG 4.7 must be to foster among the younger generations deep awareness of, and empathy regarding, the origins and consequences of inequalities at all scales; and therefore we strongly suggest a discussion on the relationship between rising social inequality, sustainability, climate change, and economic growth, thus bringing in SDGs 10 and 13 (Diffenbaugh & Burke, 2019; Taconet et al., 2020). This approach also gives students the chance to become aware of, and question, their own lifestyles and relationships to the economic system, and to speculate about what a sustainable lifestyle might look like on individual and collective bases. How we can redefine and achieve human prosperity while also respecting "planetary boundaries" (Rockström et al., 2009; O'Neill et al., 2018) and the limits to economic growth described in this paper then becomes a central question, once we have taken off our epistemological blinkers. The fact that no nation has successfully achieved human development goals without violating planetary boundaries is a sober reminder of the need – in the spirit of transformative and transdisciplinary education – to rethink, reinvent, and renegotiate taken-for-granted concepts – from endless growth to the meaning of well-being to our troubled relationship with the rest of nature.

## Notes

1 Quote from Wiedmann et al. (2020): "The world's top 10% of income earners are responsible for between 25 and 43% of environmental impact. In contrast, the world's bottom 10% income earners exert only around 3–5% of environmental impact (Teixidó-Figueras et al., 2016). These findings mean that environmental impact is to a large extent caused and driven by the world's rich citizens (Chancel and Piketty, 2016)."

2 In the literal mathematical sense.

3 Such as heavy machinery, chemicals such as pesticides, and personal gadgets and appliances.

4 It may perhaps be more appropriately termed the first wildlife extermination event given human attribution.

5 This formulation is compatible with the notion of planetary boundaries, which represents human well-being within the biophysical constraints of the planet. (Rockstrom et al., 2009).

6 In other words, of simultaneously achieving SDGs 8 and 12.

7 It is important to note that a certain minimum amount of energy will always be necessary to power the car. This means that in a car with a hypothetical engine efficiency of 100%, none of the energy in the fuel gets wasted as heat to overcome friction in the different moving parts. Thus, "20 to 40% engine efficiency" signifies that between 60 and 80% of the energy in the fuel is wasted as largely unusable heat, instead of being used to power the motion and electronics of the car.

8 The Second Law of Thermodynamics, in particular, sets a universal upper bound to the efficiency of any process governed by thermodynamics (called heat engines) called the Carnot efficiency (Carnot developed this idea in 1824). Heat engines constitute a wide class of processes that convert thermal energy into mechanical energy.

9 The Second Law's being "universal" is to be understood as meaning that it applies not just to currently known technology, but also to any yet-to-be-discovered technology as well.

10 In "combined-cycle heat engines" (GE, 2017).

11 In the specific thought experiment of the car engine, resource refers to the fuel powering the car. While this example may be very specific, the conclusions apply to any process that uses energy to operate, including energy from renewable sources. Even if the car were powered by solar panels, its engine efficiency would be limited, since friction persists in wasting energy as heat.

12 Raw materials extracted from the domestic territory plus all physical imports minus all physical exports. Not included in this category are the upstream raw materials related to imports and exports originating from outside of the focal economy.

13 The MF does not record the actual physical movement of materials within and among countries but, instead, enumerates the link between the beginning of a production chain (where raw materials are extracted from the natural environment) and its end (where a product or service is consumed) (Wiedmann et al., 2015).

14 At which point, physical and pragmatic limits to efficiency will most likely have set in, particularly if the increases in efficiency are exponential in the truly mathematical sense. An example of this is the exponential rise in processing power over the past five decades, which is now nearly at an end since chip components have been reduced to nearly atomic size: a miniaturization limit that cannot be breached.

## References

Alexander, S. (2015). *Sufficiency economy: Enough, for everyone, forever.* Simplicity Institute.

Alexander, S., & Gleeson, B. (2019). *Degrowth in the suburbs: A radical urban imaginary.* Berlin: Springer.

Bain, K. (2004). What makes great teachers great? *The Chronicle Review, 50*(31), B7.

Brundtland, G. (1987). *Report of the World Commission on Environment and Development: Our common future* (UN General Assembly document A/42/427). United Nations.

Ceballos, G., Ehrlich, P. R., & Dirzo, R. (2017). Biological annihilation via the ongoing sixth mass extinction signaled by vertebrate population losses and declines. *Proceedings of the National Academy of Sciences of the United States of America, 114*(30), E6089–E6096. doi: 10.1073/pnas.1704949114

Coffey, C., Espinoza Revollo, P., Harvey, R., Lawson, M., Parvez Butt, A., Piaget, K., … Thekkudan, J. (2020). *Time to care: Unpaid and underpaid care work and the global inequality crisis.* Oxfam. https://doi.org/10.21201/2020.5419

Daly, H. E. (2014). *From uneconomic growth to a steady-state economy: Advances in ecological economics.* Edward Elgar.

Daly, H. E., & Farley, J. (2003). *Ecological economics: Principles and applications* (1st ed.). Washington, DC: Island Press.

Davis, S. J., & Caldeira, K. (2010). Consumption-based accounting of CO2 emissions *Proceedings of the National Academy of Sciences of the United States of America, 107*(12), 5687–5692. doi: 10.1073/pnas.0906974107

Dharma Rao, C. V. (n.d.). *Water use efficiency.* Ministry of Environment, Forest, and Climate Change, Government of India. http://nwm.gov.in/sites/default/files/1.%20National-water-mission-%20%20%20water-use-efficiency.pdf

Diffenbaugh, N., & Burke, M. (2019). Global warming has increased global economic inequality. *Proceedings of the National Academy of Sciences of the United States of America, 116* (20) 9808-9813; DOI: 10.1073/pnas.1816020116

Do the Math. (2011). *Don't be a PV efficiency snob.* https://dothemath.ucsd.edu/2011/09/dont-be-a-pv-efficiency-snob/

Druckman, A., Chitnis, M., Sorrell, S., & Jackson, T. (2011). Missing carbon reductions? Exploring rebound and backfire effects in UK households. *Energy Policy 39,* 3572–3581.

Ekholm, B., & Rockström, J. (2019). *Digital technology can cut global emissions by 15%. Here's how.* https://www.weforum.org/agenda/2019/01/why- digitalization-is-the-key-to-exponential-climate-action/

Ellen Macarthur Foundation. (2017). *What is a circular economy?* https://www.ellenmacarthurfoundation.org/circular-economy/concept

English, M. C., & Kitsantas, A. (2013). Supporting student self-regulated learning in problem- and project-based learning. *Interdisciplinary Journal of Problem-Based Learning, 7*(2). https://doi.org/10.7771/1541-5015.1339

European Commission. (2020). *Sustainability.* Brussels: European Commission. https://ec.europa.eu/growth/industry/sustainability/circular-economy_en

FAO [Food and Agriculture Organization]. (n.d.). *Food loss and waste.* Rome: Food and Agriculture Organization of the United Nations. http://www.fao.org/policy-support/policy-themes/food-loss-food-waste/en/

Friends of Bernie Sanders. (n.d.). *Issues: The green new deal.* Burlington, VT: Friends of Bernie Sanders. https://berniesanders.com/issues/green-new-deal/

GE. (2017). *HA technology now available at industry-first 64 percent efficiency.* https://www.ge.com/news/press-releases/ha-technology-now-available-industry-first-64-percent-efficiency

Gerber, J.-F., & Raina, R. S. (2018). Post-growth in the global south? Some reflections from India and Bhutan. *Ecological Economics, 150.* https://doi.org/10.1016/j.ecolecon.2018.02.020

Giampietro, M., & Mayumi, K. (2018). Unraveling the complexity of the Jevons Paradox: The link between innovation, efficiency, and sustainability. *Frontiers in Energy Research, 6,* 1–13. doi: 10.3389/fenrg.2018.00026

Gilboy, J. (2017). *Mercedes-AMG's F1 engine has cracked 50 percent thermal efficiency, report says.* The Drive. https://www.thedrive.com/tech/14286/mercedes-amgs-f1-engine-has-cracked-50-percent-thermal-efficiency-report-says

Harris-White, B. (2006, April 1). Poverty and capitalism. *Economic and Political Weekly, 41.*

Hawken, P. (2017). *Drawdown: The most comprehensive plan ever proposed to reverse global warming.* New York: Penguin.

Hickel, J. (2019). The contradiction of the sustainable development goals: Growth versus ecology on a finite planet. *Sustainable Development, 27,* 873–884.

Hickel, J., & Kallis, G. (2019). Is green growth possible? *New Political Economy, 25,* 469–486.

Holmes, A. (2017). *How many times can that be recycled?* Earth 911. https://earth911.com/business-policy/how-many-times-recycled/

Howard, B. C. (2018). *Recycling myths busted: What really happens to all that stuff you put in the blue bins?* National Geographic. https://www.nationalgeographic.com/environment/2018/10/5-recycling-myths-busted-plastic/

IPBES [Intergovernmental Science-Policy Platform on Biodiversity and Ecosystem Services] (2019). *Global assessment report on biodiversity and ecosystem services of the Intergovernmental Science-Policy Platform on Biodiversity and Ecosystem Services.* Bonn: IPBES Secretariat.

IPCC [Intergovernmental Panel on Climate Change]. (2018). *Global warming of 1.5°C: An IPCC Special Report on the impacts of global warming of 1.5°C above pre-industrial levels and related global greenhouse gas emission pathways, in the context of strengthening the global response to the threat of climate change, sustainable development, and efforts to eradicate poverty.* IPCC. https://www.ipcc.ch/site/assets/uploads/sites/2/2019/06/SR15_Full_Report_High_Res.pdf

Jackson, T. (2009). *Prosperity without growth: Economics for a finite planet.* London: Earthscan.

Journal of Transformative Education. (n.d.). Journal description. *Journal of Transformative Education.* https://journals.sagepub.com/description/jtd

Kallis, G., Kostakis, V., Lange, S., Muraca, B., Paulson, S., & Schmelzer, M. (2018). Research on degrowth. *Annual Review of Environment and Resources, 43*(1), 291–316. https://doi.org/10.1146/annurev-environ-102017-025941

Kothari, A., Salleh, A., Escobar, A., Demaria, F., & Acosta, A. (Eds.). (2019). *Pluriverse: A post-development dictionary.* Tulika Books.

Kwauk, C. (2020). Roadmaps to quality education in a time of climate change [Brief]. Brookings Institute. https://www.brookings.edu/wp-content/uploads/2020/02/Roadblocks-to-quality-education-in-a-time-of-climate-change-FINAL.pdf

Labour for a Green New Deal. (n.d.). *Our plan for a just, green recovery.* https://www.labourgnd.uk

Malik, A., McBain, D., Wiedmann, T. O., Lenzen, M., & Murray, J. (2019). Advancements in input-output models and indicators for consumption-based accounting. *Journal of Industrial Based Ecology, 23,* 300–312.

McBain, D., & Alsamawi, A. (2014). Quantitative accounting for social economic indicators. *Natural Resources Forum 38,* 193–202.

Mezirow, J., & Taylor, E. W. (Eds.). (2009). *Transformative learning in practice: Insights from community, workplace, and higher education.* San Francisco, CA: Jossey-Bass.

Miller, l., Brunsell, N., Mechem, D., Gans, F., Monaghan, A., Vautard, R., … Kleidon, A. (2015). Two methods for estimating limits to large-scale wind power generation. *Proceedings of the National Academy of Sciences of the United States of America, 112*(36), 11169-11174. https://doi.org/10.1073/pnas.1408251112

Nelson, A., & Timmermans, F. (2011). *Life without money: Building fair and sustainable economies.* London: Pluto Press.

Odell, V., Molthan-Hill, P., Martin, S., & Sterling, S. (2020). Transformative education to address all sustainable development goals. In W. Leal Filho, A. M. Azul, L. Brandli, P. G. Özuyar, & T. Wall (Eds.), *Quality education* (pp. 905–916). Springer. https://doi.org/10.1007/978-3-319-95870-5_106

OECD. (2011) *Resource productivity in the G8 and the OECD: A report in the framework of the Kobe 3R action plan.* Paris: OECD.

OECD. (2020). *Re-circle: Resource efficiency and circular economy.* Paris: OECD. https://www.oecd.org/environment/waste/recircle.htm

Office of Energy Efficiency and Renewable Energy. (n.d). *Where the energy goes: Gasoline vehicles.* US Department of Energy, and Environmental Protection Agency. https://www.fueleconomy.gov/feg/atv.shtml

O'Neill, D. W., Fanning, A. L., Lamb, W. F., & Steinberger, J. K. (2018). A good life for all within planetary boundaries. *Nature Sustainability, 1*(2), 88–95. https://doi.org/10.1038/s41893-018-0021-4

Our World in Data. (n.d.). Metal production over the long term, world, 1880 to 2013. https://ourworldindata.org/grapher/metal-production-long-term

Our World in Data. (2017). *Average years of schooling vs. GDP per capita, 2017.* Our World in Data. https://ourworldindata.org/grapher/average-years-of-schooling-vs-gdp-per-capita

Parrique, T., Barth, J., Briens, F., Kerschner, C., Kraus-Polk, A., Kuokkanen A., & Spangenberg, J. H.. (2019). *Decoupling debunked: Evidence and arguments against green growth as a sole strategy for sustainability.* European Environmental Bureau. https://mk0eeborgicuypctuf7e.kinstacdn.com/wp-content/uploads/2019/07/Decoupling-Debunked.pdf

Recycling Centers. (2020). *Recycling techniques.* Jefferson, IN: Recycling Centers. https://www.recyclingcenters.org/Recycling_techniques.php

Richters, O., & Siemoneit, A. (2019). Growth imperatives: Substantiating a contested concept. *Structural Change and Economic Dynamics, 51,* 126–137.

Ritchie, H. (2017a). *Fossil fuels.* Our World in Data. https://ourworldindata.org/fossil-fuels

Ritchie, H. (2017b). *Water use and stress.* Our World in Data. https://ourworldindata.org/water-use-stress

Ritchie, H. (2018). *Plastic pollution.* Our World in Data. https://ourworldindata.org/plastic-pollution

Ritchie, H., & Roser, M. (2013). *Land use.* Our World in Data. https://ourworldindata.org/land-use

Ritchie, H., & Roser, M. (2017). *$CO_2$ and greenhouse gas emissions.* Our World in Data https://ourworldindata.org/co2-and-other-greenhouse-gas-emissions

Rockström, J., Steffen, W., Noone, K., Persson, Å., Chapin, F. S. I., Lambin, E., … Foley, J. (2009). Planetary boundaries: Exploring the safe operating space for humanity. *Ecology and Society, 14*(2). https://doi.org/10.5751/ES-03180-140232

Roser, M. (2013). *Economic growth.* Our World in Data. https://ourworldindata.org/economic-growth

Roser, M., Ortiz-Ospina, E., & Ritchie, H. (2013). *Life expectancy.* Our World in Data. 'https://ourworldindata.org/life-expectancy

Roser, M., Ritchie, H., & Dadonaite, B. (2013). *Child and infant mortality.* Our World in Data. https://ourworldindata.org/child-mortality

Singh, V. (2020). *Teaching climate change in a physics classroom: Towards a transdisciplinary approach.* Ithaca, NY: Cornell University. http://arxiv.org/abs/2008.00281

Smith, R. (2016). *Green capitalism: The god that failed.* Bristol, UK: World Economics Association.

Sterling, S. (2011). Transformative learning and sustainability: Sketching the conceptual ground. *Learning and Teaching in Higher Education, 5,* 17–33.

Sustainable Development Knowledge Platform. (n.d.). *Green growth.* United Nations. https://sustainabledevelopment.un.org/index.php?menu=1447

Taconet, N., Méjean, A., & Guivarch, C. (2020). Influence of climate change impacts and mitigation costs on inequality between countries. *Climatic Change, 160,* 15–34. https://doi.org/10.1007/s10584-019-02637-w

Teixidó-Figueras, J., Steinberger, J. K., Krausmann, F., Haberl, H., Weidmann, T., Peters, G. P., … Kastner, T. (2016). International inequality of environmental pressures: Decomposition and comparative analysis. *Ecological Indicators, 62,* 163–173.

UN Statistics Division (n.d.) *Sustainable development goals overview: 12 Responsible production and consumption: Ensure sustainable production and consumption patterns.* United Nations. https://unstats.un.org/sdgs/report/2019/goal-12/

UNEP [UN Environment Programme]. (2019). *The circular economy & the sustainable management of minerals & metal resources* (Conference invitation). United Nations. https://wedocs.unep.org/bitstream/handle/20.500.11822/30844/Circular_Economy_CN.pdf?sequence=1&isAll owed=y

UNESCO. (2016). *Education for people and the planet: Creating sustainable futures for all.* UNESCO. http://uis.unesco.org/sites/default/files/documents/education-for-people-and-planet-creating-sustainable-futures-for-all-gemr-2016-en.pdf

United Nations. (2015). *Transforming our world: The 2030 agenda for sustainable development.* New York: United Nations.

van den Bergh, J. C. J. M. (2017). A third option for climate policy within potential limits to growth. *Nature Climate Change 7,* 107–112.

Victor, P. A. (2019). *Managing without growth: Slower by design, not disaster* (2d ed.). Cheltenham, UK: Edward Elgar.

von Kessel, I. (2017). *The materials that make up the phone.* Statista. https://www.statista.com/chart/10719/materials-used-in-iphone-6/

Weinzettel, J., Hertwich, E. G., Peters, G. P., Steen-Olsen, K., & Galli, A. (2013). Affluence drives the global displacement of land use. *Global Environmental Change 23*(2), 433–438. doi: 10.1016/j.gloenvcha.2012.12.010

Weissberg, R. P., Durlak, J. A., Domitrovich, C. E., & Gullotta, T. P. (Eds.). (2015). *Social and emotional learning: Past, present, and future.* In J. A. Durlak, C. E. Domitrovich, R. P. Weissberg, & T. P. Gullotta (Eds.), *Handbook of social and emotional learning: Research and practice* (pp. 3–19). New York: Guilford Press.

Wiedmann, T., Lenzen, M., Keyßer, L. T., & Steinberger, J. K. (2020). Scientists' warning on affluence. *Nature Communications, 11*(1), 1–10. doi: 10.1038/s41467-020-16941-y

Wiedmann, T. O., Schandl, H., Lenzen, M., Moran, D., Suh, S., West, J., & Kanemoto, K. (2015). The material footprint of nations. *Proceedings of the National Academy of Sciences of the United States of America, 112*(20), 6271–6276. https://doi.org/10.1073/pnas.1220362110

World Economic Forum (n.d.). *The limits of linear consumption.* https://reports.weforum.org/toward-the-circular-economy-accelerating-the-scale-up-across-global-supply-chains/the-limits-of-linear-consumption/

Yale Poorvu Center for Teaching and Learning. (n.d.). *Case-based Learning.* https://poorvucenter.yale.edu/faculty-resources/strategies-teaching/case-based-learning